\documentclass[aps,prl,twocolumn,superscriptaddress,floatfix,showpacs]{revtex4-2}

\usepackage{graphicx}
\usepackage{amsmath,amssymb,amsfonts}
\usepackage{xcolor}
\usepackage{bm}
\usepackage{booktabs}
\usepackage[colorlinks=true,linkcolor=blue,citecolor=blue,urlcolor=blue]{hyperref}

\newcommand{\WSe}{WSe$_{2}$}
\newcommand{\MoSe}{MoSe$_{2}$}

\newcommand{\MoS}{MoS$_{2}$}

\newcommand{\SiO}{SiO$_{2}$}
\newcommand{\SiN}{Si$_{3}$N$_{4}$}

\begin{document}

\title{Tuning Terahertz Optomechanics of MoS$_2$ Bilayers with Homogeneous In-plane Strain}

\author{S. Patel}
\affiliation{Department of Physics, University of Arkansas, Fayetteville, Arkansas 72701, USA}

\author{Jose D. Mella}
\affiliation{Department of Physics, University of Arkansas, Fayetteville, Arkansas 72701, USA}

\author{S. Puri}
\affiliation{Department of Physics, University of Arkansas, Fayetteville, Arkansas 72701, USA}

\author{Salvador Barraza-Lopez}
\affiliation{Department of Physics, University of Arkansas, Fayetteville, Arkansas 72701, USA}
\affiliation{Institute for Solid State Physics, University of Tokyo, Kashiwanoha, 277-8581, Chiba, Japan}

\author{H. Nakamura}
\email{hnakamur@uark.edu}
\affiliation{Department of Physics, University of Arkansas, Fayetteville, Arkansas 72701, USA}

\date{\today}

\begin{abstract}
\noindent{}Homogeneous in-plane biaxial tensile strain strengthens the out-of-plane van der Waals (vdW) interaction in \MoS\ bilayers (BLs) and can be used to fine-tune their terahertz (THz) oscillations. Using ultralow-frequency Raman spectroscopy on hexagonal (2H) and rhombohedral (2R) stacked BLs, we observe a hardening of the interlayer breathing modes originating from a strain-induced Poisson contraction of the vdW separation between the layers characterized by an effective out-of-plane Poisson's ratio of $\nu_\mathrm{eff} \approx 0.19\text{--}0.24$. Strikingly, this geometric contraction drives the system into a highly repulsive regime of the intermolecular potential, corresponding to a Gr\"uneisen parameter of $\gamma \approx 14\text{--}20$. This value surpasses even the `giant' one reported for phosphorene, establishing these van der Waals BLs as highly tunable nonlinear mechanical platforms that can be addressed at the THz regime without external pressure knobs.
\end{abstract}

\maketitle

Weather monitoring, airport security, and high-bandwidth (6G) communications can all benefit from electromagnetic oscillators at the THz range \cite{thz_roadmap,book}. While van der Waals (vdW) materials are actively explored to directly create or detect THz radiation \cite{sharma2025probing,bhuyan2026enhanced}, their Raman-active THz phonon vibrations provide a powerful parallel platform for optomechanical control, with potential applications ranging from microwave acoustic filters to acousto-optic modulators \cite{yoon_terahertz_2024}. Through coherent excitation via stimulated Raman effects, these THz lattice vibrations can dynamically modulate macroscopic optical properties, including transmission \cite{jeong2016coherent}, reflection \cite{kim2024ultrafast}, and nonlinear frequency conversion \cite{maehrlein2017terahertz}. Furthermore, at the quantum limit, these localized vibrations hold promise for single-phonon manipulation, as demonstrated using THz phonons in diamond \cite{lee2011entangling}.

Strain permits tailoring the electronic and optical landscapes of two-dimensional (2D) materials \cite{roldan2015strain,amorim2016novel,naumis2017electronic, peng2020strain, naumis2024mechanical} by inducing gauge fields \cite{pachecosanjuan2014graphene}, shifting excitonic emission energies \cite{conley2013bandgap,he2016strain}, and by enhancing electron mobility \cite{yang2024biaxial,datye2022strain}. However, despite the extensive study of intralayer responses, the optomechanics of the interlayer vdW coupling has only begun to be explored \cite{du2025interlayer}. A fundamental question when strain is created in-plane is the extent to which the mechanical deformation within the plane propagates across monolayers (MLs).

Understanding this cross-coupling is critical for designing vdW heterostructures, where the subtle interplay between MLs dictates phenomena ranging from moir\'e excitons and hyperlubricity to superconductivity and topological phases \cite{bian2022strong, tran2019evidence, cao2018unconventional, sharpe2019emergent}. In this direction, Du and coworkers studied few-ML MoS$_2$ under isotropic compressive strain \cite{du2025interlayer}.  Here, we consider anisotropic coupling between in-plane and out-of-plane strain instead.

\begin{figure}[tb]
    \centering
    \includegraphics[width=0.48\textwidth]{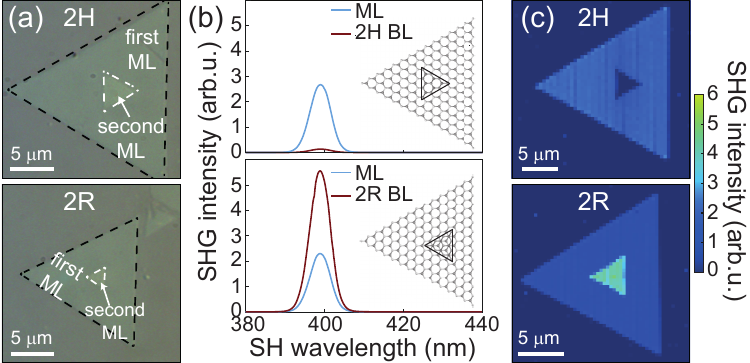}
    \caption{(a) Optical contrast images of 2H and 2R MoS$_2$ BLs grown on a \SiO{} substrate. (b) SHG spectra for ML, 2H and 2R BLs. Insets are ball and stick models. (c) SHG intensity maps in ML and BL regions.}
    \label{fig:1}
\end{figure}

We synthesized \MoS\ BLs under biaxial tensile strain  \cite{patel2024biaxial} on \SiO\ [Fig.~\ref{fig:1}(a)] and \SiN\ substrates (section S1 in Ref.~\cite{supmat}); MLs develop triangular shapes. Bilayer flakes are created by the successive growth of triangular MLs; smaller triangles around the centers of larger ones in Fig.~\ref{fig:1}(a) are showcased within white dashed lines. The top ML shows different orientations depending on the BL phase: it rotates by 60$^{\circ}$ relative to the lower ML in the 2H phase, but it forms a concentric structure with parallel edges in the 2R BL \cite{marmolejo2022slippery}. Our first-principles calculations \cite{kresse1996efficiency,optPBE,PBE,grimme2010consistent} (section S2 in Ref.~\cite{supmat}) show an energy difference of only $\sim$0.7 meV per unit cell between the two phases, resulting in a similar likelihood for either configuration to be created.

A single-location second-harmonic generation (SHG, section S3 in Ref.~\cite{supmat}) spectrum within ML sections is displayed in blue on Fig.~\ref{fig:1}(b); peaks are centered at half the wavelength of the incident laser. Due to its center of inversion, the nominal SHG from a 2H BL is zero; the small nonzero SHG intensity seen in the top panel is likely due to a small breaking of inversion symmetry within a substrate-material-air interface. In contrast, the SHG signal on the 2R BL (dark red curve in the bottom panel) is more than twice that of the ML one. Chalcogen atoms form zigzag edges, and the edge conformation of 2H and 2R BLs was recreated with ball-and-stick models in the insets of Fig.~\ref{fig:1}(b). The presence of a center of inversion was spatially probed by SHG in Fig.~\ref{fig:1}(c), which greatly increases the contrast of both 2H and 2R BL sections when compared to that obtained with an optical microscope [Fig.~\ref{fig:1}(a)].

\begin{figure}[tb]
    \centering
    \includegraphics[width=0.48\textwidth]{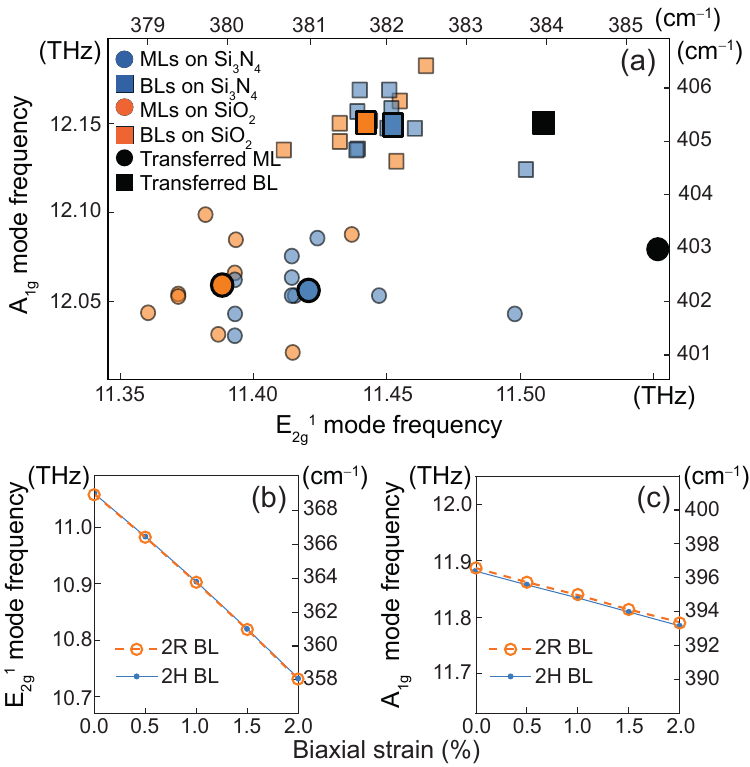}
    \caption{(a) Correlation between $A_{1g}$ and $E_{2g}^1$ peak positions for MLs (small circles) and 2H BLs (small squares), including average values (large circles and squares). The black circle (square) corresponds to a sample where the strain was released after transferring a \MoS\ ML (BL) onto another \SiO\ substrate. (b) $E_{2g}^1$ and (c) $A_{1g}$ mode frequencies from DFT calculations for 2H (orange) and 2R (light blue) phases, respectively. Both modes red-shift under in-plane biaxial strain.}
    \label{fig:3}
\end{figure}

The response of the high-frequency (in-plane intralayer $E_{2g}^1$ and out-of-plane $A_{1g}$ vibrations) Raman \cite{itoh_reliable_2020,supmat} intralayer modes is the primary gauge for quantifying isotropic pressure \cite{du2025interlayer} and strain \cite{chakraborty2012symmetry,lloyd2016band,rao2019spectroscopic,patel2024biaxial,gontijo2025resonant} in layered materials. While uniaxial strain breaks hexagonal symmetry and splits the $E_{2g}^1$ degeneracy \cite{wang2013raman}, biaxial strain preserves that symmetry, leading to a systematic redshift of the $E_{2g}^1$ mode with increasing tension \cite{lloyd2016band}.

Biaxial strain is determined from the thermal expansion mismatch between the TMDC and the growth substrate~\cite{ahn2017strain,patel2024biaxial}, estimated as $\mathrm{Strain(\%)} = \left(\alpha_{\mathrm{MoS_2}}-\alpha_\mathrm{sub}\right) \Delta T \times 100$, where $\alpha$ denotes the thermal expansion coefficients and $\Delta T$ is the difference between growth ($T_\mathrm{growth}$) and room ($T_\mathrm{room}$) temperatures \cite{ding2015thermal,roy1989very,sinha1978thermal}. Therefore, our \MoS\ samples are subjected to a tensile biaxial strain of $+0.36\%$ on \SiN\ and $+0.72\%$ on \SiO\ (section S4 in Ref.~\cite{supmat}).

\begin{figure}[tb]
    \centering
    \includegraphics[width=0.48\textwidth]{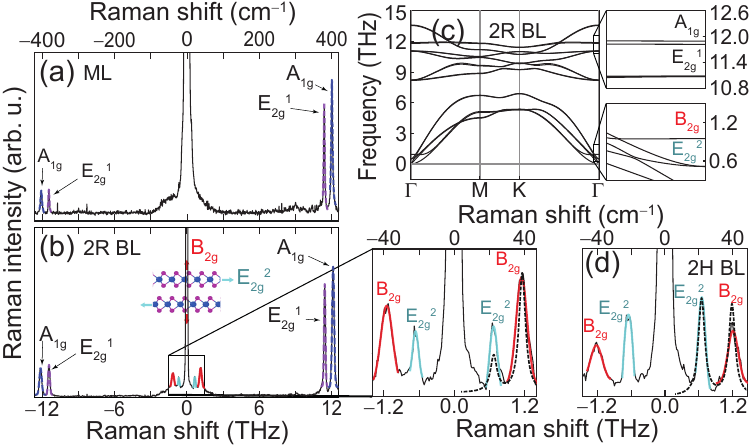}
    \caption{(a) Raman spectrum of a \MoS\ ML on \SiO. (b) Raman spectrum of a 2R BL on \SiO. The magnified view between $-1.5$ and 1.5 THz shows signatures of shear ($E_{2g}^2$) and breathing ($B_{2g}$) modes absent in the monolayer. (c) Strain-free phonon dispersion for a 2R bilayer. Raman-active high- and low-frequency modes are highlighted on insets. (d) Low-frequency Raman spectrum of 2H phase on \SiO. Theoretical Raman spectra are shown with the dashed curves in the zoom-in of panel (b) and in panel (d) for $\boldsymbol{g}_i = \boldsymbol{g}_s = (1,0,0)^T$.}
    \label{fig:2}
\end{figure}

The substrate-dependent evolution of the $A_{1g}$ and $E_{2g}^1$ modes for MoS$_2$ MLs (circles) and BLs (squares) is shown in Fig.~\ref{fig:3}(a). Frequencies of individual flakes were obtained by fitting a Lorentzian profile to each peak. They appear in lighter-colored symbols, from which average values (larger, opaque symbols) were calculated. The $A_{1g}$ mode has a negligible substrate dependence \cite{supmat} with average values of 12.05 THz and 12.14 THz (402~cm$^{-1}$ and 405~cm$^{-1}$) for MLs and BLs, respectively. In contrast, the $E_{2g}^1$ mode redshifts, reflecting a greater tensile strain when on \SiO\ than on \SiN\ (section S5 in Ref.~\cite{supmat}). (Alternative fits based on Gaussians gave almost identical frequencies; see Table SI in  Ref.~\cite{supmat}. The Raman spectra of Fig.~\ref{fig:2} and Figs.~S4, S6 and S8 on Ref.~\cite{supmat} were fitted against Gaussians.)

Frequency renormalization (and hence tensile strain) is further evidenced by comparisons against strain-free control \MoS\ ML and BL samples transferred onto \SiO/Si \cite{haley2021heated} (see Section S6 and Fig.~S5 in Ref.~\cite{supmat}). To start with, the Raman frequencies of control samples are consistent with previous measurements \cite{zhang_raman_2013,yang_temperature-dependent_2024} (see Table~\ref{tab:1}).

In the relaxed control sample, the $E_{2g}^1$ mode {\em blueshifts} by 0.16~THz (5.3~cm$^{-1}$) for the ML and 0.06~THz (2~cm$^{-1}$) for the BL, relative to the as-grown samples on \SiO, while the $A_{1g}$ mode shifts by only $\sim$0.03~THz ($\sim$1~cm$^{-1}$) [Fig.~\ref{fig:3}(a)].

\begin{table}[tb]
    \centering
    \caption{Comparison of intralayer ($A_{1g}$  and $E_{2g}^1$) modes for unstrained MoS$_2$ MLs and BLs against the literature.}\label{tab:1}
    \begin{tabular}{cccccc}
        \toprule
        \toprule
        Reference & $E_{2g}^1$ & $A_{1g}$ & Reference  & $E_{2g}^1$ & $A_{1g}$\\
        (ML) &(cm$^{-1}$)&(cm$^{-1}$)&(BL)&(cm$^{-1})$&(cm$^{-1}$)\\
\midrule
        Ref.~\cite{zhang_raman_2013} & 384 & 403 & Ref.~\cite{zhang_raman_2013}      &385 & 408     \\
        Ref.~\cite{yang_temperature-dependent_2024}       & 384 & 403 & Ref.~\cite{yang_temperature-dependent_2024}   &384 & 404\\
        \textbf{This work} & {\bf 384} & {\bf 405} & \textbf{This work}&{\bf 385} & {\bf 403}    \\ \bottomrule
        \bottomrule
    \end{tabular}%
\end{table}

The strain trends are reproduced by theoretical results [Figs.~\ref{fig:3}(b,c)], where the sensitivity of the $E_{2g}^1$ mode to strain is found to be about three times that of the $A_{1g}$ mode.  Using the measured frequency difference between as-grown and strain-released 2H-BLs, we extract a strain-tuning rate (strain-induced red shift) for the \textit{intralayer} $E_{2g}^1$ mode equal to $-0.10$~THz$/\%$ ($-$3.3~cm$^{-1}/\%$) which aligns with our theoretical estimate of $-0.15$~THz/\% ($-$5~cm$^{-1}/\%$) reasonably well. Fig.~\ref{fig:2}(a) is a prototypical Raman spectrum contributing to the ML statistics on Fig.~\ref{fig:3}(a).

While high-energy phonons map the in-plane stress field, they offer limited insight into the evolution of the interlayer coupling itself, and we next turn to the low-frequency Raman spectra near 1 THz (33.3~cm$^{-1}$) in Fig.~\ref{fig:2}(b)---where the rigid-layer Shear ($E_{2g}^2$) and Breathing ($B_{2g}$) modes reside---to probe the vertical interaction among MLs. Phonon dispersions for strain-free 2R [Fig.~\ref{fig:2}(c)] and 2H \cite{supmat} BLs were calculated as well \cite{phonopy-phono3py-JPCM,phonopy-phono3py-JPSJ}.

The mechanical response of interlayer modes to strain is not completely understood. Previous work showed uniaxial strain splitting the degenerate $E_{2g}^2$ modes \cite{lee2017strain}, yet the behavior of the $B_{2g}$ mode---which corresponds to the out-of-plane oscillation and is a direct proxy for the interlayer distance---was studied under hydrostatic compressive strain \cite{du2025interlayer}; not biaxial tensile strain.

The zoom-in on Fig.~\ref{fig:2}(b) and Fig.~\ref{fig:2}(d) contrast low-frequency spectra for the 2H and 2R BLs, respectively. The Raman amplitude of the $E_{2g}^2$ mode (0.67~THz or 22.3~cm$^{-1}$) exceeds that of the $B_{2g}$ mode (1.20~THz or 40~cm$^{-1}$) in the 2H BL (see section S7 in Ref.~\cite{supmat}, too). In contrast, this intensity ratio is inverted in the 2R BL, where the $B_{2g}$ mode has a larger amplitude. This reversal in amplitude reflects the different symmetries of the two stacking orders.

The difference in amplitudes is confirmed by Raman calculations of the low-frequency modes. The differential cross section (proportional to the Raman intensity) far from resonance is given by \cite{liang_first-principles_2014,porezag_infrared_1996,bagheri_high-throughput_2023,supmat}:

\begin{equation}
    \frac{d\sigma_j}{d\Omega}=\frac{\omega_s^4}{c^4}\frac{\hbar(n_j+1)}{2\omega_j} I^{\text{R}}_j,
\end{equation}
where $c$ is the speed of light. Energy conservation imposes $\omega_s=\omega_i - \omega_j$, with $\omega_s$ the frequency of the scattered light, $\omega_j$ the frequency of the $j-$th phonon mode of the crystal, and $\omega_i$ the incoming scattered light frequency. The Bose factor of the $j-$th phonon mode is $n_j=(e^{\hbar \omega_j/k_B T}-1)^{-1}$, where $k_B$, $\hbar$, and $T$ are the Boltzmann constant, the reduced Planck constant and temperature, respectively. $I_j^\text{R} = |\boldsymbol{g}_s^T R_j \boldsymbol{g}_i|^2$ is the Raman activity of the $j$-th phonon mode, where $\boldsymbol{g}_i$ and $\boldsymbol{g}_s$ are the polarization vectors of the incoming and scattered light. $R_j$ is the Raman tensor for the vibrational mode $j$. A detailed description is provided in Refs.~\cite{bagheri_high-throughput_2023}. The dashed curves in the zoomed-in panel of Fig.~\ref{fig:2}(b) and in Fig.~\ref{fig:2}(d) are room-temperature theoretical Raman spectra for the 2R and 2H bilayers, respectively. It demonstrates good agreement with the experimental results.

The $E_{2g}^2$-mode degeneracy, protected by the $C_3$ symmetry of the hexagonal lattice, serves as a sensitive probe for symmetry-breaking perturbations such as uniaxial strain \cite{lee2017strain}. The lack of $E_{2g}^2$-mode splitting in our BL samples, as well as the isotropic intensity for the polarization-resolved SHG of 2R-\MoS\ BL (Fig.~S3 in Ref.~\cite{supmat}), are both consistent with the uniform, biaxial nature of strain~\cite{patel2024biaxial,puri2024substrate}.

\begin{figure}[tb]
    \centering
    \includegraphics[width=0.48\textwidth]{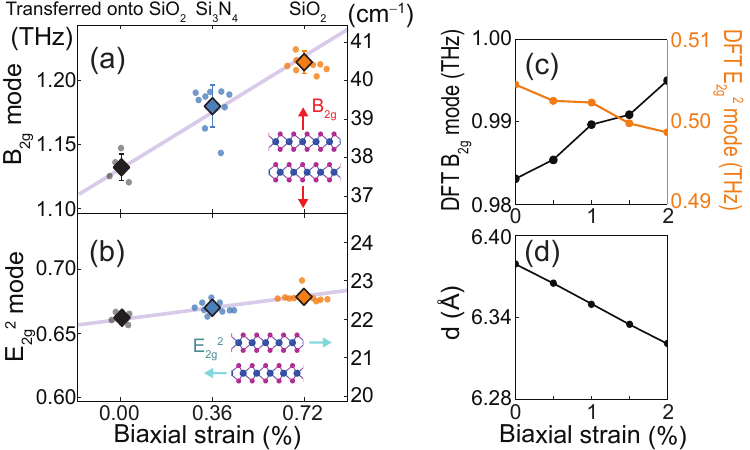}
    \caption{Experimental evolution of the (a) $B_{2g}$ and (b) $E_{2g}^2$ frequencies for 2H-stacked $\mathrm{MoS_{2}}$ BLs under varying biaxial tensile strain. (c) DFT calculations of the breathing (black) and shear (orange) mode frequencies for 2H \MoS. (d) Interlayer distance $d$ {\em vs.}~in-plane biaxial strain.}
    \label{fig:4}
\end{figure}

We next correlate the response of the interlayer coupling to the strain created during growth. We display the measured low-frequency modes for 2H (Figs.~\ref{fig:4}(a,b)) and 2R (Fig.~S7 and section S8 in Ref.~\cite{supmat}) BLs on both growth substrates. We include 2H BL samples for which strain was released by transferring the BL onto a \SiO\ substrate (section S8 in Ref.~\cite{supmat}). The $E_{2g}^2$ mode undergoes a slight blueshift under tensile strain (Fig.~\ref{fig:4}(b); and Ref.~\cite{supmat}). However, the most striking feature is the response of {\em the $B_{2g}$ mode}, which {\em undergoes a significant hardening with increasing strain}: the average $B_{2g}$-mode frequency increases from  1.13~THz (37.7~cm$^{-1}$) for 0\% strain (transferred) to 1.21~THz (40.4~cm$^{-1}$) for +0.72\% strain (\SiO), yielding a substantial experimental strain-tuning rate of $+0.11$~THz/\% ($+3.7$~cm$^{-1}/\%$). Thus, we have demonstrated that biaxial strain created by growth can be used \textit{in lieu} of isotropic strain to tune the optomechanical response of MoS$_2$ BLs at the $\sim 1$ THz oscillator range. This response remains in place as strain is built-in during growth.

This hardening of the out-of-plane vibration under in-plane tension is counter-intuitive for standard harmonic oscillators and it arises from the Poisson effect. As the \MoS\ BL is stretched biaxially, the interlayer spacing contracts to conserve volume, steepening the confinement potential. We corroborate this mechanism using first-principles calculations, which reproduce the monotonic hardening of the $B_{2g}$ mode (Fig.~\ref{fig:4}(c)) and confirm the reduction in ML spacing $d$ (Fig.~\ref{fig:4}(d)).

The interlayer phonon frequencies and their renormalization due to strain are sensitive to the approximation for the vdW interaction used in calculations \cite{supmat}. However, the Poisson compression (vertical contraction, $\Delta d$) driving the $B_{2g}$-mode hardening was robust irrespective of exchange-correlation functional. We also note that while the theory captures the qualitative trend, the predicted magnitude of the shift ($\sim$0.01~THz/\% or $\sim$0.3~cm$^{-1}/\%$) is significantly smaller than in experiments. This quantitative discrepancy highlights the difficulty standard functionals have in fully capturing the strain-sensitivity of the vdW potential anharmonicity. Nevertheless, the geometric response—the vertical contraction of the layer spacing—is robust across different functionals (see Fig.~S2 in Ref.~\cite{supmat}).

%{\color{red} Biaxial strain extends the accessible frequency range for coherent acoustic phonon generation in TMDCs. Since the highest frequencies are found in bilayers and decrease with increasing number of layers, strain provides an additional tuning mechanism to push the phonon generation range further into the THz regime.}

We rely on a geometric response to quantify the effective Poisson's ratio ($\nu_\mathrm{eff}$) of BLs (section S9 in Ref.~\cite{supmat}). Using the relation for isotropic biaxial strain assuming the BL behaves as a continuum elastic medium \cite{landau1986theory}:
\begin{equation}
    \nu_\mathrm{eff} = \frac{-\varepsilon_{\perp} / \varepsilon_{\parallel}}{2 - \varepsilon_{\perp} / \varepsilon_{\parallel}} \text{,}
\end{equation}
where $\varepsilon_{\parallel}$ is the applied in-plane strain and $\varepsilon_{\perp}$ is the induced vertical strain derived from the theoretical vdWs gap contraction shown in Figure~\ref{fig:4}d (section S10 in Ref.~\cite{supmat}). This calculation yields $\nu_\mathrm{eff}= 0.19-0.24$, depending on the exchange-correlation approximation. As shown in Table~\ref{tab:poisson_comparison}, this value is notably lower than those reported for other three-dimensional materials as well as for bulk \MoS\ ($\nu_\mathrm{eff} \approx 0.30$) \cite{feldman1976elastic}. We note that the literature value for bulk \MoS\ was derived indirectly from linear compressibility measurements, while theoretical estimates for the relevant elastic constant ($C_{13}$) have shown significant scatter (3--39 GPa) depending on the treatment of vdW interactions \cite{peelaers2014elastic}. Our result aligns closely with modern hybrid-functional DFT (HSE06-D2) results, which predict a lower bulk Poisson's ratio of $\nu_\mathrm{eff} \approx 0.17$ \cite{peelaers2014elastic}.

One notable discrepancy between the experiment and DFT results is how much the interlayer phonon hardens in response to the Poisson compression, which can be described by the Gr\"uneisen parameter. To the best of our knowledge, the Gr\"uneisen parameter for the interlayer breathing mode in a few-ML transition metal dichalcogenides (TMDs) has not been previously determined experimentally. While intralayer modes ($E_{2g}^1$, $A_{1g}$) in \MoS, \MoSe, and \WSe\ typically exhibit Gr\"uneisen parameters near unity ($\gamma \approx 1$) \cite{dadgar2018strain, huang2016low}, our measurement yields a significantly larger value. We estimate the out-of-plane Gr\"uneisen parameter ($\gamma_{out}$) using the definition:
\begin{equation}
    \gamma_{out} = -\frac{1}{\omega_0} \frac{\partial \omega}{\partial \varepsilon_{\perp}} \text{,}
\end{equation}
where $\omega_0$ is the zero-strain frequency (1.13~THz) and $\partial \omega / \partial \varepsilon_{\perp}$ is the rate of frequency shift with respect to vertical strain. By combining our experimental tuning rate ($\partial \omega / \partial \varepsilon_{\parallel} \approx 0.11$~THz$/\%$) with the theoretical Poisson contraction ratio ($\varepsilon_{\perp}/\varepsilon_{\parallel} \approx -0.47$ to $-0.65$), we obtain $\gamma_{out} \approx 14\text{--}20$. This value exceeds even the `giant' $\gamma$ reported for phosphorene ($\gamma \approx 8.6$) \cite{cai2015giant}, highlighting the extreme sensitivity of the curvature of vdW potential to the geometrical compression.

\begin{table}[tb]
    \centering
    \caption{Poisson's ratios ($\nu$) for representative 3D semiconductors, perovskite oxides, and layered materials. The value for the \MoS\ BL is estimated using $d$ [Fig.~\ref{fig:4}(d)].}
    \label{tab:poisson_comparison}
    \resizebox{\columnwidth}{!}{%
    \begin{tabular}{llcc}
        \toprule
        \toprule
        Material Class & Material & Poisson's Ratio ($\nu$) & Ref. \\ \midrule
        Group IV & Silicon (Si) & 0.28 & \cite{wortman1965young} \\

        & Germanium (Ge) & 0.27 & \cite{wortman1965young} \\ \midrule
        III-V & GaAs & 0.31 & \cite{blakemore1982semiconducting} \\ \midrule
        Perovskite & SrTiO$_3$ & 0.25 & \cite{bell1963elastic} \\
         & BaTiO$_3$ & 0.31 & \cite{berlincourt1958elastic} \\ \midrule
        Layered  & Graphite ($\nu_{out}$) & 0.29 & \cite{blakslee1970lastic} \\ \midrule
        Layered TMDs & \MoS\ (Bulk, Exp.) & $\sim$0.30 & \cite{feldman1976elastic} \\

        & \MoS\ (Bulk, Theo.) & $\sim$0.17 & \cite{peelaers2014elastic} \\ \midrule
        \textbf{This work} & \textbf{\MoS\ BL} & \textbf{0.19-0.24} & -- \\ \bottomrule
        \bottomrule
    \end{tabular}%
    }
\end{table}
%37.8~\text{cm}=1.13 THz
%2.5~\text{cm}=0.07 THz

Our findings redefine the vdW spacing in BL systems as a dynamic, highly tunable degree of freedom at the THz oscillation regime. We established that the soft vertical coupling acts as a sensitive amplifier of in-plane mechanical deformation. The resulting stiffening of the interlayer bond implies a direct modulation of the electronic interlayer hopping parameter ($t_{\perp}$). This electromechanical coupling provides a physical basis for dynamically manipulating the layer hybridization of excitons, thereby controlling their lifetimes and valley coherence~\cite{alexeev2019resonantly, rivera2018interlayer}. Furthermore, as the field advances toward twisted heterostructures, this effect offers a deterministic knob to tune the bandwidth of moiré flat bands without changing the twist angle~\cite{bi2019designing, li2025strain}, paving the way for strain-tunable correlated phases and optomechanics.

\begin{acknowledgments}
We thank R.~Rodriguez, M.~Marking, K.~Reynolds, H.~Churchill, and P.~Kumar. This work is supported by the Office of the Secretary of Defense for Research and Engineering (Award No.~FA9550-23-1-0500) and by the NSF's Q-AMASE-i program (Award No.~DMR-1906383). Calculations were performed at the Arkansas HPCC (NSF Award No.~OAC-2346752).
\end{acknowledgments}

%\bibliography{biblio_May22}

%%%%%%%%%% Merge with supplemental materials %%%%%%%%%%
\clearpage
\onecolumngrid
%\appendix

% Reset counters for figures, tables, etc.
\setcounter{figure}{0}
\setcounter{table}{0}
\setcounter{equation}{0}
\setcounter{page}{1}
\setcounter{secnumdepth}{2} % number sections + subsections (use 3 for subsubsections)

% Prefix "S" to labels
\renewcommand{\thesection}{S\arabic{section}}
\renewcommand{\thesubsection}{S\arabic{section}.\arabic{subsection}}
\renewcommand{\thefigure}{S\arabic{figure}}
\renewcommand{\thetable}{S\arabic{table}}
\renewcommand{\theequation}{S\arabic{equation}}

%%%%%%%%%% End Merge setup %%%%%%%%%%

% --- SM title block with correct affiliations ---
\begin{center}
\vspace*{-1.0em}
\begin{minipage}{0.99\textwidth}
\centering

{\large\bfseries Supplemental Material for}\\[2pt]
{\large\bfseries Tuning Terahertz Optomechanics of MoS$_2$ Bilayers\\ with Homogeneous In-plane Strain}\\[8pt]

{\normalsize
S. Patel\textsuperscript{1},
Jose D. Mella\textsuperscript{1},
S. Puri\textsuperscript{1},
Salvador Barraza-Lopez\textsuperscript{1,2},
H. Nakamura\textsuperscript{1}\\[4pt]

\textsuperscript{1}\textit{Department of Physics, University of Arkansas, Fayetteville, AR 72701, USA}\\
\textsuperscript{2}\textit{Institute for Solid State Physics, University of Tokyo, Kashiwa, 277-8581, Chiba, Japan}\\[4pt]

%\href{mailto:hnakamur@uark.edu}{hnakamur@uark.edu}
}

\end{minipage}
\vspace*{0.5em}
\end{center}
% --- end SM title block ---

% ---------------------------------------------------------
% SECTIONS
% ---------------------------------------------------------

\section{Growth of $MoS_2$ bilayers}
\MoS\ bilayer films were grown using a home-built physical vapor deposition (PVD) system at a growth temperature of 1200$^\circ$C using pure \MoS\ powder as the source material~\cite{patel2024biaxial}. The high growth temperature enables a large tensile strain originating from thermal expansion mismatch between amorphous substrates and the 2D material~\cite{patel2024biaxial}. The PVD system is equipped with two mass flow controllers on both sides of a quartz tube (1 inch diameter) that regulate the Ar flow to prevent unwanted deposition. Following a growth time of 5-10 seconds, we rapidly quenched the temperature by sliding the furnace away from the location of the powder and the substrate. Silicon substrates with either a 90 nm custom-grown oxide, or a 70 nm silicon nitride, enhance the optical contrast of grown bilayer films.

%\clearpage
\section{\textit{Ab initio} calculations}

DFT calculations were performed with the {\em Vienna Ab Initio Simulation Package} (VASP) \cite{kresse1996efficiency} using PAW pseudopotentials and the optPBE-vdW density functional \cite{optPBE}. We used a 18$\times$18$\times$1 $k$-point mesh, an energy cutoff of 400 eV, and an electronic energy convergence criteria of $10^{-8}$ eV. A force convergence of $10^{-6} ~\text{eV/\AA}$ was set for structural relaxations. All calculations were performed with an out-of-plane lattice constant of $c\sim 60$ \AA~to reduce self-interaction among periodic copies.\\

The 2R phase is 0.7 meV/u.c. higher in energy than the 2H one after structural relaxation. The lattice parameter for both 2H and 2R phases  is 3.20 \AA. The equilibrium interlayer sulfur distance for the 2H (3R) stacked bilayer is 3.23 \AA{} (3.22 \AA{}). The monolayer thickness, defined as the out-of-plane S–S distance within a single \MoS\ monolayer, is 3.15 \AA{} for both stacking configurations. The distance between monolayers was defined as $z_1 - z_2$, where $z_{1},z_2$ denote the geometric centers of the top and bottom monolayers, respectively.\\

The applied strain is $a=a_0(1+\varepsilon)$, where $a$ and $a_0$ denote the lattice parameter with and without strain and $\varepsilon$ is the (dimensionless) strain. The lattice parameter was modified by a desired percentage, and atomic positions were relaxed while keeping the lattice vectors fixed.\\

Phonons were calculated within the harmonic approximation using the Phonopy package \cite{phonopy-phono3py-JPCM,phonopy-phono3py-JPSJ} after a full structural relaxation at each value of strain $\varepsilon$. The phonon dispersion was calculated for a 5$\times$5$\times$1 supercell with a displacement of 0.05 \AA~and a 3$\times$3$\times$1 $k$-point mesh.\\

%Figure~\ref{fig:S8} contains results obtained from PBE exchange-correlation functional combined with the empirical vdW correction proposed by Grimme \textit{et al}. \cite{PBE,grimme2010consistent} (DFT-D3) with zero-damping function.

\begin{figure}[h!]
    \centering
    \includegraphics[width=0.96\linewidth]{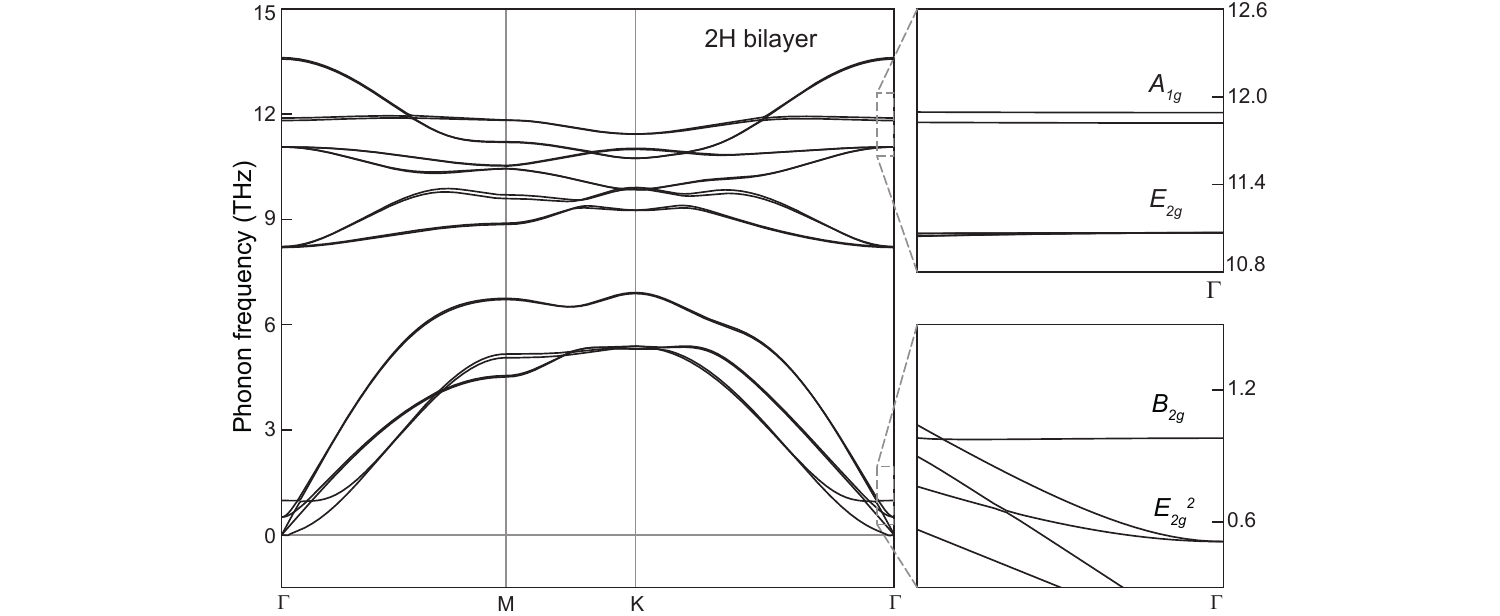}
    \caption{Phonon dispersion of a strain-free 2H-\MoS\ bilayer with the optPBE-vdW approximation. Relevant modes around the $\Gamma-$point are highlighted on the panels to the right.}
    \label{fig:S1}
\end{figure}

The phonon dispersions for the unstrained 2R and 2H BLs are shown in Fig.~3(c) of the main text and in Fig.~\ref{fig:S1}, respectively, while the response to strain of the vibrational modes is shown in Fig.~2(b,c) and Fig.~4(c,d) of the main text.\\

\begin{figure}[h]
    \centering
    \includegraphics[width=0.96\linewidth]{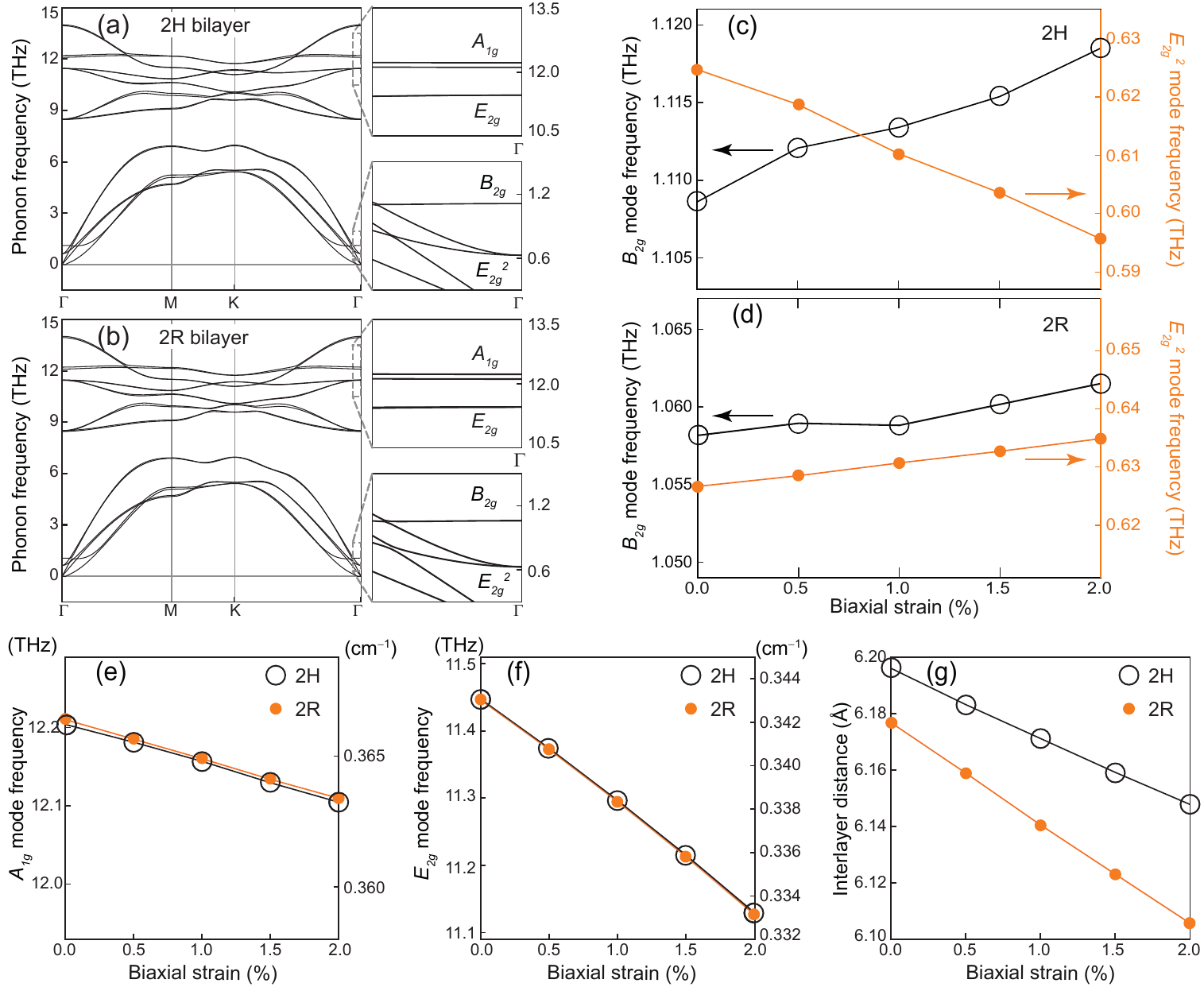}
    \caption{Vibrational properties of 2H and 2R \MoS\ bilayers using the DFT-D3 approximation. Phonon dispersion of (a) 2H and (b) 2R bilayers with a magnified view at the $\Gamma-$point highlighting the $B_{2g}$ and $E^2_{2g}$ modes, as well as the intralayer phonon branches. Interlayer low frequency modes for (c) 2H and (d) 2R bilayers. (The vertical scale is 0.6 cm$^{-1}$ for $B_{2g}$ mode and 16 cm$^{-1}$ for $E^2_{2g}$ mode in panels (c) and (d).) (e) $A_{1g}$  and (f) $E_{2g}$  modes obtained for 2H (orange) and 2R (light blue) phases as a function of in-plane biaxial strain. (g) Interlayer distance for both phases under biaxial strain.}
    \label{fig:S8}
\end{figure}

Setting IVDW = 11 as an input, we also included van der Waals (vdW) effects through semi-empirical DFT-D3 corrections within the PBE functional \cite{PBE,grimme2010consistent}. The lattice parameters in that case were 3.16 \AA{} and 3.17 A{} for the 2H and 2R phases, respectively. The monolayer thickness was 3.13 \AA{} for both phases, and the interlayer sulfur distance is 3.06 \AA{} (2H) and 3.05 \AA{} (2R). Fig.~\ref{fig:S8} shows the vibrational spectra within this approximation: calculations with the optPBE-vdW and semiempricial DFT-D3 van der Waals corrections have the same qualitative trends.\\

Raman spectra were calculated for 2H and 2R BLs. The Raman tensor for the $j-$th vibrational mode is:
\begin{equation}
    R_{jlm}=\frac{V_c}{4\pi}\frac{\partial \chi_{lm}}{\partial \xi_j}.
\end{equation}
Here, $\chi_{lm}$ is the macroscopic electronic susceptibility tensor and $V_c$ is the volume of the unit cell. $\xi_j$ is a generalized coordinate that characterizes the collective displacement of all atoms along the mass-scaled eigenvector $\boldsymbol{e}'_j$. If $\boldsymbol{e}_j$ is the eigenvector obtained from the diagonalization of the dynamical matrix, the relation between the entries of $\boldsymbol{e}'_j$ and $\boldsymbol{e}_j$ is given by $e'_{jk}(\kappa) = e_{jk}(\kappa)/\sqrt{M_{\kappa}}$, where $e_{jk}(\kappa)$ and $e'_{jk}(\kappa)$ denote the entry for the atom $\kappa$ with mass $M_{\kappa}$ along the Cartesian direction $k$. The directional derivative is calculated as:
\begin{equation}
    \frac{\partial \chi_{lm}}{\partial \xi_j}=\frac{\chi_{lm}(\mathbf{R}_0+h\hat{\boldsymbol{e}}_j')-\chi_{lm}(\mathbf{R}_0-h\hat{\boldsymbol{e}}_j')}{2h}|\hat{\boldsymbol{e}}_j'|,
\end{equation}
with $\textbf{R}_0$ a vector with the equilibrium positions of all the atoms. Use of the normalized vector $\hat{\boldsymbol{e}}_j'$ guarantees having a consistent displacement $h$. The change in $\chi_{lm}$ is evaluated using the macroscopic dielectric constant $\varepsilon_{lm}=1+4\pi\chi_{lm}$, as provided by VASP.

%\clearpage

\section{Second Harmonic Generation and Raman spectroscopies}
\begin{figure}[h!]
    \centering
    \includegraphics[width=0.96\linewidth]{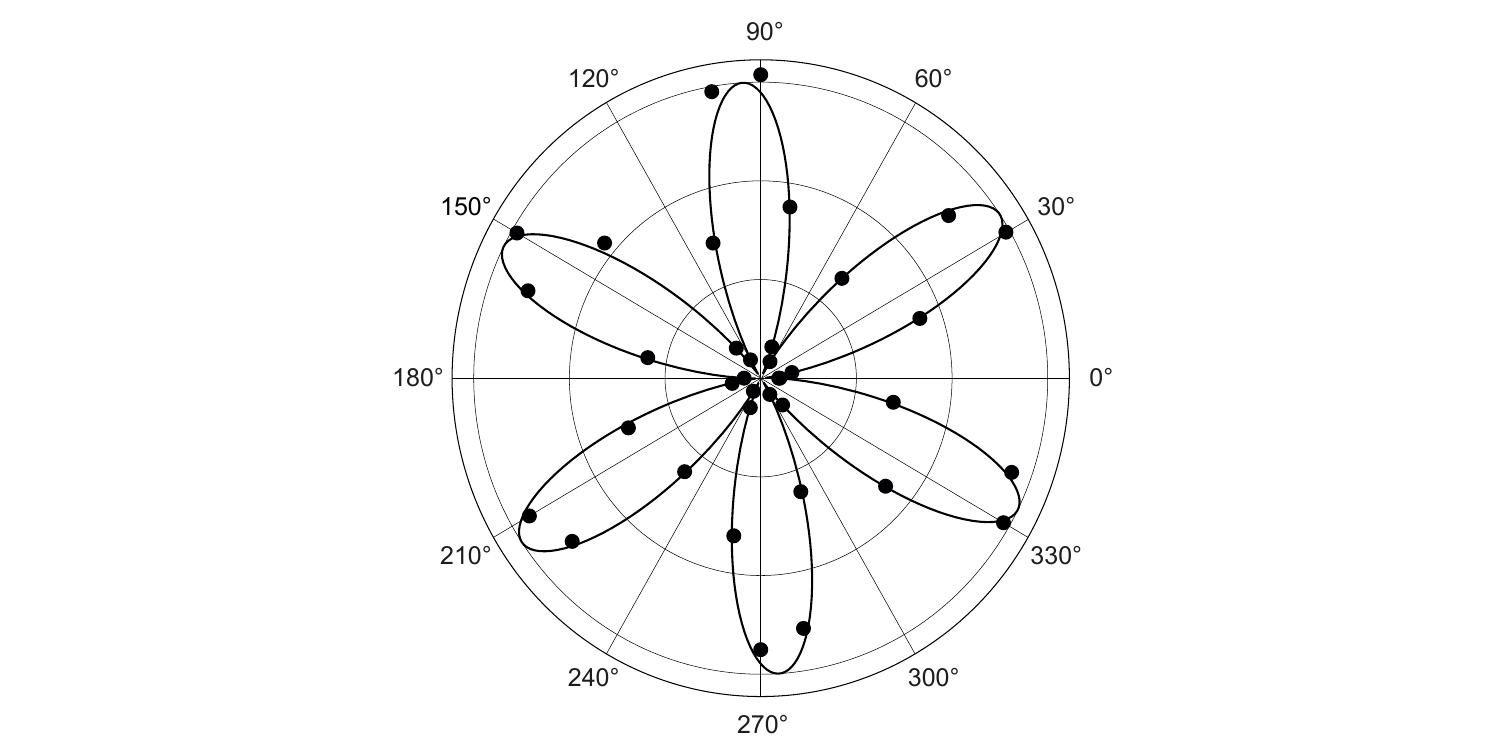}
    \caption{The symmetric six-lobed pattern reflects that the in-plane crystal symmetry is well preserved ({\em i.e.}, there are no directionally-preferred distortions of the \MoS\ bilayer), confirming biaxial strain.}
    \label{fig:S3}
\end{figure}

Details of the second harmonic generation (SHG) setup are as follows: a Ti:sapphire fs oscillator (Tsunami, Spectra-Physics) with an 80 MHz repetition frequency was used as an excitation laser, which delivered a temporal pulse width of $\sim$100 fs with $\sim$1 mW at the position of the sample. The reflected SHG was observed through filters that rejected the fundamental beam, and the angular dependence was taken by rotating the sample. Motorized XY stages allowed for a spatially-resolved SHG mapping by translating the sample. All Raman and SHG measurements were performed in air at room temperature. In addition to the SHG intensity maps and spectra shown in the main text, the angular SHG of the 2R-MoS$_2$ bilayer is shown in Fig.~\ref{fig:S3}; its sixfold symmetry confirms that the strain truly is biaxial.

Raman measurements were carried out using a home-built system equipped with a 532 nm excitation laser. An objective lens with 50× (NA = 0.55) magnification was used to focus the light on the sample, and the laser power on the sample was below 2 mW. We used two narrow-line OD4 Bragg type filters (Optigrate) to reject the laser line, which allowed us to measure Raman shifts down to $\sim$20 cm$^{-1}$ for both Stokes and anti-Stokes peaks. Raman spectra were calibrated using the optical phonon peak of the silicon substrate at 520.2 cm$^{-1}$; a well-known standard \cite{itoh_reliable_2020}. We show a comparison of the Lorentzian and Gaussian fitting for the interlayer modes ($E_{2g}^2$ and $B_{2g}$) in Table~\ref{tab:comparison2}.\\

\begin{table}[h!]
    \centering
    \caption{Comparison of intralayer and interlayer frequency modes between Lorentzian and Gaussian fitting. Both 2H and 2R \MoS\ on \SiO\ and \SiN\ substrates were considered. These frequencies were extracted from peak maxima and averaged over nine flakes.}\label{tab:comparison2}
    \begin{tabular}{ccccc}
        \toprule
        \toprule
        Structure         & \multicolumn{2}{c}{$E_{2g}^2$ (cm$^{-1}$)} & \multicolumn{2}{c}{$B_{2g}$ (cm$^{-1}$)}  \\
                         & Lorentzian     & Gaussian      & Lorentzian  & Gaussian            \\ \midrule
        2H \MoS\ on \SiO & 22.58          & 22.60         & 40.51       & 40.49    \\
      2H \MoS\ on \SiN & 22.30          & 22.34         & 39.35       & 39.49    \\
        2R \MoS\ on \SiO & 22.71          & 22.71         & 39.14       & 39.08   \\
        2R \MoS\ on \SiN & 22.03          & 22.08         & 38.75       & 38.71  \\ \bottomrule
        \bottomrule
    \end{tabular}%
\end{table}

\section{Estimation of strain due to thermal expansion mismatch}
The biaxial strain due to mismatched thermal expansion coefficients of the transition metal dichalcogenide (TMD) bilayer and the underlying substrate is \cite{ahn2017strain,patel2024biaxial}:
\begin{equation}
    \%\,\text{strain} = (\alpha_{\mathrm{TMD}} - \alpha_{\mathrm{sub}})\cdot \Delta T \cdot 100,
\end{equation}
where $\alpha_{\mathrm{TMD}}$ is the thermal expansion coefficient of \MoS, $\alpha_{\mathrm{sub}}$ is that of \SiN\ or \SiO, and $\Delta T = T_{\text{growth}} - T_{\text{room}}$. This expression is approximate because it uses average (constant) thermal expansion coefficients over the relevant temperature range. We employed the following values: $\alpha_{\text{MoS}_2} = 6.6 \times 10^{-6}\,\text{K}^{-1}$\cite{ding2015thermal}, $\alpha_{\text{SiO}_2} = 5.0 \times 10^{-7}\,\text{K}^{-1}$\cite{roy1989very}, and $\alpha_{\text{Si}_3\text{N}_4} = 3.5 \times 10^{-6}\,\text{K}^{-1}$\cite{sinha1978thermal}. Using $T_{\text{room}} = 25^{\circ}\mathrm{C}$ and $T_{\text{growth}} = 1200^{\circ}\mathrm{C}$, we estimate tensile strains of $+0.36\%$ for \MoS\ on \SiN, and $+0.72\%$ for \MoS\ on \SiO. These results indicate that thermal-expansion mismatch can induce substantial tensile strain in TMD films grown at elevated temperatures, with the largest effects occurring on \SiO\ substrates.

\section{Substrate-dependent high frequency Raman modes of 2H-$MoS_2$ bilayers}

\begin{figure}[h]
    \centering
    \includegraphics[width=0.96\linewidth]{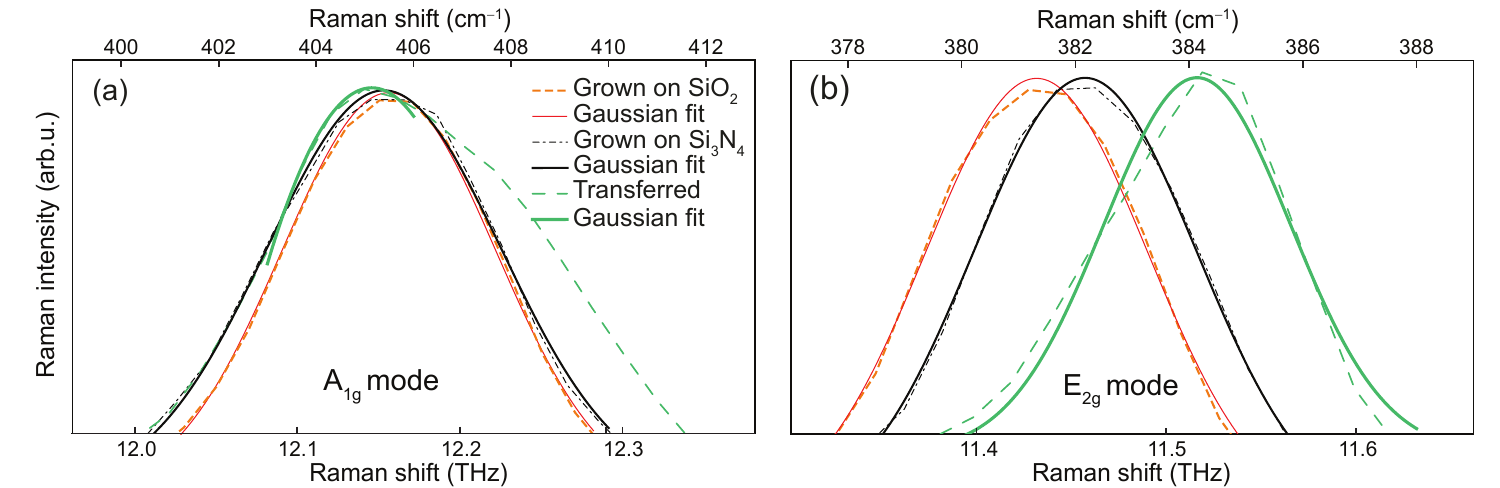}
    \caption{(a) Raman spectra showing the out-of-plane $A_{1g}$ mode of as-grown (\SiO\ and \SiN) and transferred \MoS\ bilayers. (b) Representative Raman spectra showing the in-plane (\textit{intralayer}) $E_{2g}$ mode of as-grown (\SiO\ and \SiN) and transferred \MoS\ bilayers.}
    \label{fig:S4}
\end{figure}

Fig.~\ref{fig:S4}(a) is a representative Raman spectra showing the out-of-plane $A_{1g}$ mode of as-grown (\SiO\ and \SiN) and transferred \MoS\ bilayers, whose peaks are fitted with Gaussian profiles. The $A_{1g}$ mode exhibits only subtle shifts in frequency with substrate variations, indicating that the $A_{1g}$ phonon mode is not significantly affected by biaxial tensile strain. In panel (b), a representative Raman spectra showing the in-plane (\textit{intralayer}) $E_{2g}$ mode of as-grown (\SiO\ and \SiN) and transferred \MoS\ bilayers, whose peaks are fitted with Gaussian profiles. The $E_{2g}$ peaks show a sizable redshift with increasing tensile strain.
% Fig S4: Multi-panel -> 0.8

%\clearpage
\section{Transfer of PVD-grown $MoS_2$ bilayers}
\begin{figure}[h!]
    \centering
    \includegraphics[width=0.96\linewidth]{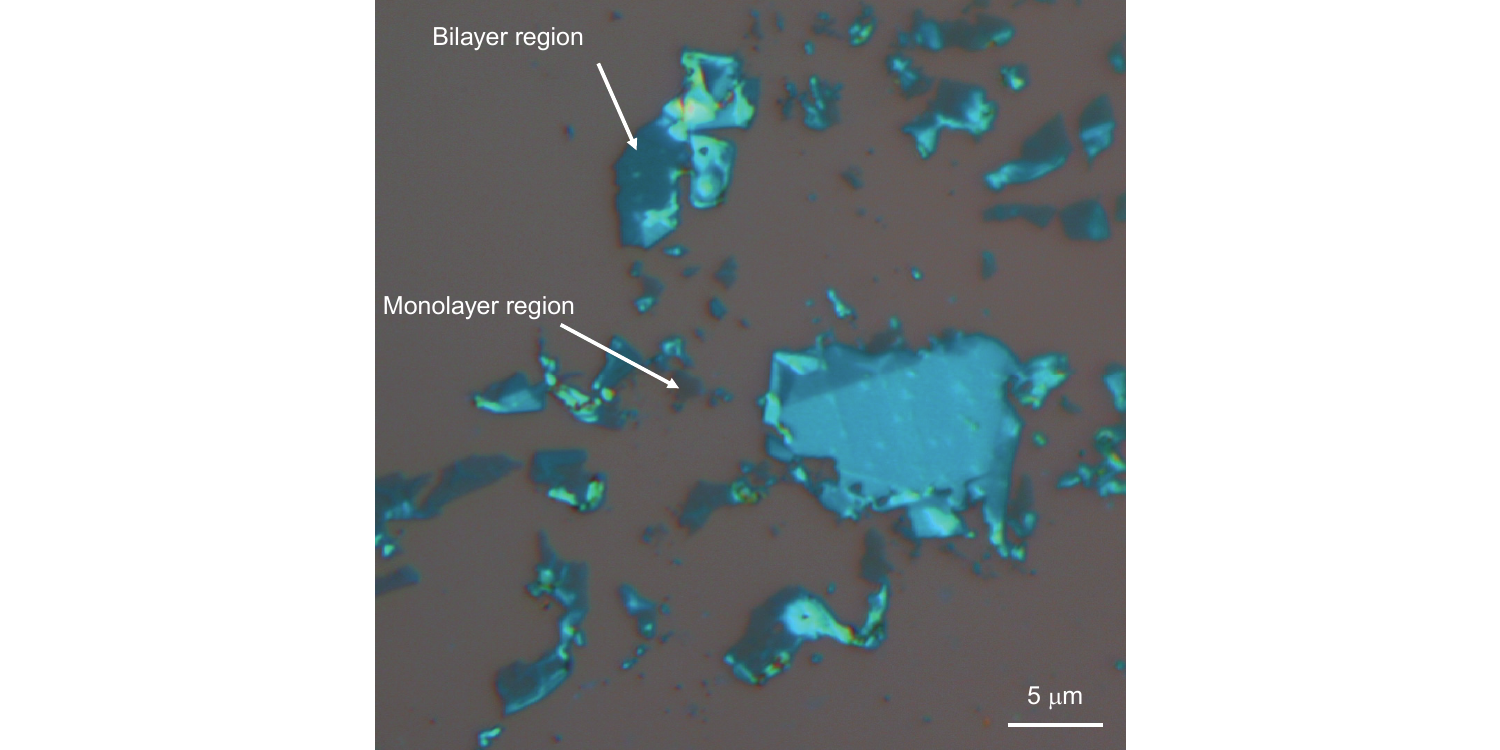}
    \caption{Optical microscopy image of PVD–grown \MoS\ flakes transferred onto a \SiO/Si substrate. Different contrast helps distinguishing monolayer and bilayer regions.}
    \label{fig:S5}
\end{figure}

The nail polish transfer method \cite{haley2021heated} was used to transfer PVD-grown \MoS\ bilayer islands on \SiO/Si to another \SiO/Si substrate to release strain. First, we placed a small square-cut polydimethylsiloxane (PDMS) on the edge of a glass slide, which we then covered with scotch tape. We put a drop of nail polish on the PDMS and then annealed the slide at 90$^\circ$C for 5 minutes. Next, we pressed the nail polish stamp over the flake that needed to be transferred from the growth substrate and heated the substrate to 80$^\circ$C. After that, we allowed the system to cool down to room temperature, and then we lifted the nail polish stamp to pick up the flake from the growth substrate. Finally, we pressed the nail polish stamp onto a fresh Si substrate with 90 nm of thermally grown \SiO. The substrate is heated to 120$^\circ$C. After the heating, the nail polish stamp is pulled up slowly, leaving \MoS\ bilayer islands along with nail polish on the substrate. Acetone and isopropyl alcohol are used to remove the nail polish. Fig.~\ref{fig:S5} shows the optical micrograph of the transferred MoS$_2$ flakes.

%\clearpage
\section{Raman spectra of shear and breathing modes of 2H-$MoS_2$ bilayers on {$SiO_2$} and {$Si_3N_4$} substrates}
\begin{figure}[h!]
    \centering
    \includegraphics[width=0.96\linewidth]{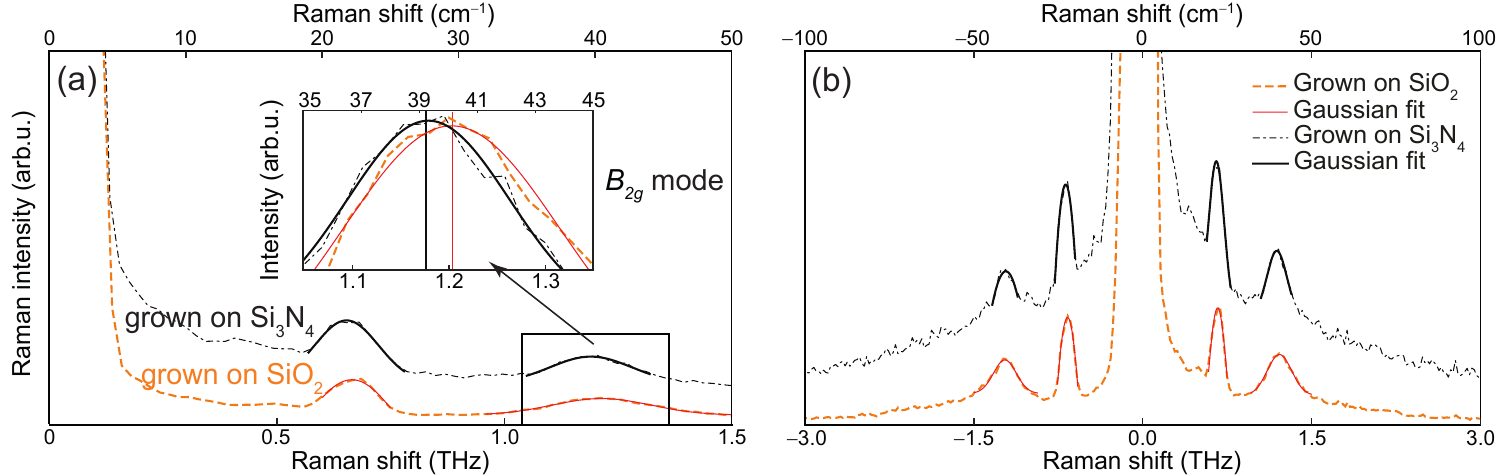}
    \caption{(a) Low-frequency Raman spectra of \MoS\ bilayers grown on \SiO\ and \SiN\ substrates. The inset highlights the $B_{2g}$ mode, and fitting with Gaussian profiles. (b) A representative Raman spectra showing Stokes and anti-Stokes peaks.}
    \label{fig:S2}
\end{figure}

 We present low-frequency Raman spectra of \MoS\ grown on \SiO\ and \SiN\ substrates in Fig.~\ref{fig:S2}, highlighting the shear ($E^2_{2g}$) and layer-breathing ($B_{2g}$) modes. The Stokes and anti-Stokes peaks are shown with Gaussian fits. We emphasize the shear ($E^2_{2g}$) and layer-breathing ($B_{2g}$) modes.
% Fig S2: Multi-panel -> 0.8

%\clearpage

\section{Interlayer vibration in $MoS_2$ bilayers under strain}

\begin{figure}[h]
    \centering
    \includegraphics[width=0.96\linewidth]{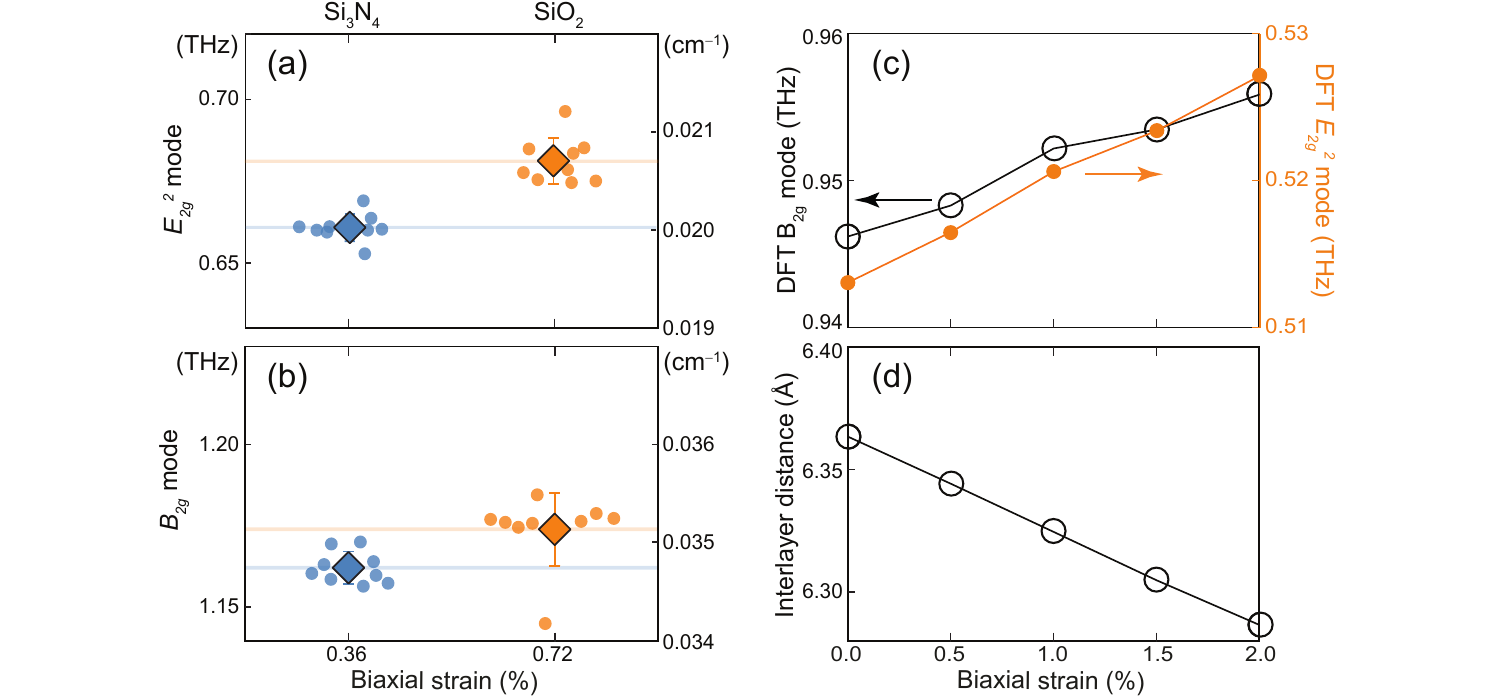}
    \caption{(a) $E^2_{2g}$ and (b) $B_{2g}$ mode frequencies for 2R-\MoS\ BL as a function of biaxial strain created by the substrate. (c) DFT calculations of the breathing (black) and shear (orange) mode frequencies for 2R-\MoS\ BLs under biaxial tensile strain. (d) DFT calculation of the interlayer distance between the two monolayers for 2R-\MoS\ BL under biaxial strain.}
    \label{fig:S6}
\end{figure}

Figs.~\ref{fig:S6}(a,b) present the strain response of 2R-\MoS\ samples grown on Si$_3$N$_4$ and SiO$_2$. Fig.~\ref{fig:S6}(c) is a calculation of these modes under biaxial strain using DFT; theory and experiment follow the same trend. Vibrational frequencies range around 0.030~THz (0.15~THz) for the $E_{2g}^{2}$ ($B_{2g}$) mode, in good agreement with the theoretical results (Fig.~\ref{fig:S6}(c)). The interlayer distance, included in Fig.~\ref{fig:S6}(d), decreases with increasing strain, consistent with the Poisson effect.

% Fig S6: Multi-panel -> 0.8

%\clearpage

\begin{figure}[h]
    \centering
    \includegraphics[width=0.96\linewidth]{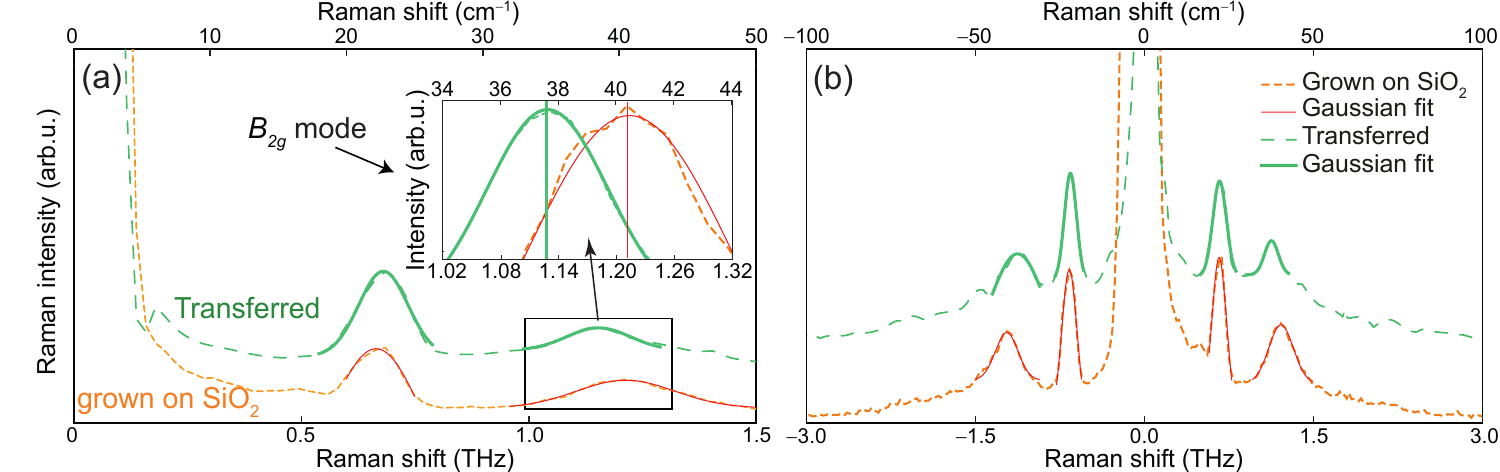}
    \caption{ (a) Representative low-frequency Raman spectra of as-grown and transferred 2H bilayer \MoS\ on \SiO, showing the $E^2_{2g}$ and $B_{2g}$ modes. The inset highlights the $B_{2g}$ mode fitted with Gaussian profiles. (b) A wider-range Raman spectra showing Stokes and anti-Stokes peaks.}
    \label{fig:S7}
\end{figure}

Fig.~\ref{fig:S7} shows a low-frequency Raman spectrum of a strain-released 2H \MoS{} bilayer on a \SiO. A considerable shift is shown as an inset of Fig.~\ref{fig:S7}(a). Fig.~\ref{fig:S7}(b) shows a corresponding zoomed-out view that includes both Stokes and anti-Stokes peaks.

% Fig S7: Multi-panel -> 0.8

\section{Derivation of Effective Poisson's Ratio}
To quantify the vertical coupling strength using a single effective parameter, we assume that the bilayer behaves as a continuum elastic medium \cite{landau1986theory}. Under in-plane stress ($\sigma_{zz}=0$) and equi-biaxial loading ($\sigma_{xx}=\sigma_{yy}=\sigma_{\parallel}$), the strains are given by:
\begin{align}
    \varepsilon_{\parallel} &= \frac{1}{Y} [\sigma_{\parallel} - \nu_\mathrm{eff} (\sigma_{\parallel} + 0)] = \frac{\sigma_{\parallel}}{Y} (1 - \nu_\mathrm{eff}),\text{ and} \\
    \varepsilon_{\perp} &= \frac{1}{Y} [0 - \nu_\mathrm{eff} (\sigma_{\parallel} + \sigma_{\parallel})] = -\frac{2\nu_\mathrm{eff}\sigma_{\parallel}}{Y} \text{.}
\end{align}
Taking the ratio of the vertical strain to the in-plane strain eliminates the Young modulus $Y$ and stress $\sigma_{\parallel}$:
\begin{equation}
    \frac{\varepsilon_{\perp}}{\varepsilon_{\parallel}} = \frac{-2\nu_\mathrm{eff}}{1 - \nu_\mathrm{eff}}\text{.}
\end{equation}
Solving this expression for $\nu_\mathrm{eff}$ yields the effective Poisson's ratio in terms of the geometric strain contraction:
\begin{equation}
    \nu_\mathrm{eff} = \frac{-\varepsilon_{\perp} / \varepsilon_{\parallel}}{2 - (\varepsilon_{\perp} / \varepsilon_{\parallel})}\text{.}
    \label{eq:nu_eff}
\end{equation}

\section{Estimating the vertical/in-plane strain ratio}
For a hexagonal crystal system (in-plane isotropic with the $c$-axis along $z$), the stress in the out-of-plane direction ($\sigma_{zz}$) is related to the strains by the stiffness tensor component:
\begin{equation}
    \sigma_{zz} = C_{13}\varepsilon_{xx} + C_{13}\varepsilon_{yy} + C_{33}\varepsilon_{zz}\text{.}
\end{equation}
In our experiment, the bilayer is subjected to equi-biaxial in-plane tension ($\varepsilon_{xx} = \varepsilon_{yy} = \varepsilon_{\parallel}$) and the vertical direction is free to relax ($\sigma_{zz} = 0$). Substituting these conditions into the equation yields:
\begin{equation}
    0 = 2 C_{13}\varepsilon_{\parallel} + C_{33}\varepsilon_{\perp} \text{.}
\end{equation}
Rearranging this terms provides the theoretical ratio between the induced vertical compression ($\varepsilon_{\perp}$) and the applied in-plane strain ($\varepsilon_{\parallel}$):
\begin{equation}
    \frac{\varepsilon_{\perp}}{\varepsilon_{\parallel}} = -2 \frac{C_{13}}{C_{33}} \text{.}
    \label{eq:strain_ratio}
\end{equation}
This relation was used to estimate the effective Poisson ratio from literature in Table I of the main text. %\cite{liang_first-principles_2014,porezag_infrared_1996,bagheri_high-throughput_2023}.

%testing page \cite{yang_temperature-dependent_2024,ling_role_2014,zhang_raman_2013}

%\bibliography{biblio_5_19_26}


\end{document}